\documentclass[12pt]{article}
\usepackage[dvips]{graphicx}
\def\v1{\vspace{1cm}}
\def\be{\begin{equation}}
\def\ee{\end{equation}}
\def\bc{\begin{center}}
\def\ec{\end{center}}

\def\vh{\varphi}

\newcommand{\bea}{\begin{eqnarray}}
\newcommand{\eea}{\end{eqnarray}}
\renewcommand{\tanh}{\rm th}

\begin{document}
\def\thebibliography#1{\section*{}\list
{\arabic{enumi}.}{\settowidth\labelwidth{#1.}\leftmargin\labelwidth
\advance\leftmargin\labelsep
\usecounter{enumi}}
\def\newblock{\hskip .11em plus .33em minus -.07em}
\sloppy
\sfcode`\.=1000\relax}
\let\endthebibliography=\endlist
\renewcommand{\thefootnote}{\arabic{footnote})}
\large \baselineskip=24pt \thispagestyle{empty}
\begin{center}
{\Large \bf Cosmological Production of Vector Bosons and Cosmic
Microwave Background Radiation
}\\[3mm]
D.B. Blaschke, S.I. Vinitsky, A.A. Gusev, V.N.
Pervushin*\footnote{ E-mail:\,\, pervush@thsun1.jinr.ru}, and \\
D.V. Proskurin \\

{\it Joint Institute for Nuclear Research, Dubna, Moscow region, 141980 Russia\\
}
\end{center}

\thispagestyle{empty}

\bigskip
\noindent \noindent{\bf Abstract} -- The intensive cosmological
creation of vector $W, ~Z$- bosons in the cosmological model with
the relative units
 is considered. Field theoretical models  are studied, which
predict that the CMB radiation and the baryon matter in the
universe can be products of decay and annihilation processes of
these primordial bosons.

\bigskip

\setcounter{page}{1}

\centerline {\it 1. INTRODUCTION}

\vspace{5mm}

Is modern theory able to explain the origin of observed matter in
the Universe by its cosmological production from a vacuum [1-11]?
 As is well known,
the answer to this question is associated with the problem of
particle creation in the vicinity of a cosmological singularity.
Thus far, it has been common practice to assume that the number of
product pairs is by far insufficient for explaining the total
amount of observed matter \cite{grib80}.

We recall that the cosmological creation of massive particles is
calculated by going over to conformal variables \cite{grib80},
  for which the limit of zero scale factor
(point of a cosmic singularity) means the vanishing of masses.
Vector bosons are the only particles of the Standard Model that
have a singularity at zero mass \cite{wen,hp}.
    In this limit, the normalization of the
wave function for massive vector bosons is singular in
mass~\cite{wen,hp}.
 The absence of the massless limit in the
theory of massive vector bosons is well known~\cite{sf}. In
calculations in the lowest order of perturbation theory this leads
to a divergence of the number of product longitudinal
bosons~\cite{grib80,ppgc}.

  There are two opinions concerning the removal
 of this singularity. In \cite{grib80,S} the divergence of the
number of particles is removed by means of a standard
renormalization of the gravitational constant. However, it is also
indicated in the monograph of Grib et al. \cite{grib80} that the
number of product particles is determined by the imaginary part of
loop Feynman diagrams; since, in quantum field theory, it is the
real parts of these diagrams that are subjected to
renormalization. This means that the above divergence of the
number of particles does not belong to the class of divergences in
quantum field theory that are removed by means of a conventional
renormalization of
    physical quantities. Indeed, the physical origin of this
divergence is that the problem of a cosmological creation of
particles from a vacuum is treated within an idealized
formulation. The point is that the quantum production of particles
in a finite volume for a system featuring interaction and exchange
effects may lead to a set of Bose particles having a specific
statistical distribution with respect to energy such that it is
able to ensure the convergence of the respective integral of the
momentum distribution.

In the present study, we analyze physical conditions and models
for which the number of product vector bosons maybe quite
sufficient for explaining the origin of matter in the Universe.
Such cosmological models include conformal cosmology
 \cite{039}, where
conformal quantities of the general theory of relativity and of
the Standard Model are defined as observables \cite{ps} for which
there are relative reference units of intervals.

 The ensuing exposition is organized as follows.
Section 2 is devoted to discussing various versions of the
formulation of the Cauchy problem for the cosmological production
of vector particles in field theory. In Section 3 we study
possible implications of such a production in the context of
validating the temperature of cosmic microwave background
radiation within the Standard Model, the baryon-antibaryon
asymmetry of the Universe \cite{sufn,ufn}, and a small
contribution of visible baryon matter \cite{fuk} to the evolution
of the Universe. In the Conclusion we discuss the results obtained
by calculating the composition of matter in the Universe within
the Standard Model.

\vspace{10mm}

\centerline {\it 2. PROBLEM OF COSMOLOGICAL PARTICLE CREATION}

\centerline {\it 2.1. Theory}

\vspace{5mm}

Let us consider cosmological particle creation in the conformally
invariant version of the general theory of relativity [9,21-24].
We have
\be\label{GR:CI}%
S_{\rm tot}[w|F]=S_{\rm D}[w|e,Q]+S_{\rm SM}[y_hw|f,e], \ee
 where,
for the action of the general relativity we take the
Penrose-Chernikov-Tagirov action functional for a scalar field
(dilaton) $w$,
\be\label{GR}%
S_{\rm D}[w|e,Q]=
\int d^4x\left[|e|w^2\left(\partial_\mu Q\partial^\mu Q-\frac{{R(e)}}{{6}}\right)+  w\partial_\mu\left(|e|
\partial^{\mu} \right)w\right],
\ee%
 in the space specified by the interval
 \be\label{ut1}
 ds^2=\left(e_{\underline \lambda \mu}dx^\mu\right)^2=
 \left(e_{\underline 0 \mu}dx^\mu\right)^2-\left(e_{\underline i
 \mu}dx^\mu\right)^2. \ee
 Here  $e_{\underline \lambda \mu}$ is Fock's vierbein, $R(e)$ is  the
 curvature, and $Q$ is an additional field that does not interact
  with matter
  \cite{ppgc} and which yields  the observed regime of
  cosmological evolution. In the Standard Model action functional
  featuring the set of fields $f$, the Higgs mass
  $M_{\mbox{\tiny Higgs}}$ is replaced by the dilaton multiplied
 by a constant
 $y_h \sim 10^{-17}$, $(y_h w)$.
 The theory specified by action functional
 (\ref{GR:CI}) is invariant under conformal
 transformations, including
 scale transformations for the set of
 all fields $[w|F]$ with the transformation parameter $\Omega$,
\be \label{conf} {}^{(n)}F_{\Omega}={}^{(n)}F \times
(\Omega)^{n},~~~~~~w_{\Omega}=\frac{w}{\Omega}, \ee where $(n)$ is
a conformal weight. This invariance indicates that the action
functional (\ref{GR:CI}) involves an extra degree of freedom.

\vspace{25mm}

\centerline {\it 2.2. Absolute Variables}

\vspace{5mm}

It is common practice to assume \cite{kl} that the action of
general theory of relativity and the Standard Model arises from
the action functional (\ref{GR:CI})
  as a
  consequence of choosing the {\it ``absolute''} variables as
\be \label{cca}
{}^{(n)}F_{(a)}={}^{(n)}F \times (w/\vh_0)^{n},~~
w_{(a)}(x^0,x^i)=\vh_0,
\ee
 with the result that the dilaton $w(x^0,x^i)$  is replaced
 the parameter
 $\vh_0$ that is related to the Planck mass
 by the equation $\vh_0=M_{\rm
Pl}(3/8\pi)^{1/2}$ and which did not appear
 in the original action  functional (\ref{GR:CI}).
 Upon the spontaneous scale-invariance breaking associated
 with this, the symmetry of the action functional (\ref{GR:CI})
 under
   the
  transformations in (\ref{conf}) becomes the symmetry of the physical
  variables in (\ref{cca}), which are invariant under the same
  scale transformations in (\ref{conf}). Owing to the above spontaneous breakdown of
scale invariance, the extra degree of freedom characterized by a
negative probability is removed from the action functional in
(\ref{GR:CI}), but, instead, the dimensional
 {\it ``absolute''}
 Planck mass parameter
   $M_{\rm Pl}$ appears in
the equations of motion. This parameter specifies initial data
concerning the emergence of the Universe in the so-called Planck
era. In the theory involving such a spontaneous breakdown of
symmetry, the homogeneous approximation of the metric, \be
\label{gr2} e_{(a)\underline{0}\mu}dx^\mu=dt,~~~
e_{(a)\underline{i}\mu}dx^\mu=a(t)dx_{\underline{i}}, \ee for
variables in (\ref{cca}) leads to  standard cosmological models
including, the inflationary model \cite{linde}, where the initial
data of the Planck era are considered as the fundamental
quantities of the equations of motion.

Within this approach, there arise problems of cosmological initial
data, the horizon, time and energy, homogeneity,singularity,and
the quantum wave function for the Universe, and attempts are made
to solve these problems at the level of the homogeneous
approximation via the inflationary expansion of absolute
space~\cite{linde}.

In \cite{pp,bpp}, some arguments are adduced that indicate that,
in all probability, all these problems, including the emergence of
the Planck era, stem from an incorrect formulation of a
spontaneous breakdown of the symmetry of (\ref{cca}) in
eliminating degrees of freedom of negative probability from the
theory specified by the action functional in (\ref{GR:CI})

 We recall that, within a gauge theory, a formulation where
 all degrees of freedom that are characterized by a negative
 probability are removed prior to quantizing the theory
 being considered is referred to as a ``fundamental method''
 \cite{sch2,pol,mpl}, in contrast to a ``heuristic method''
 \cite{fp1}, where all degrees of freedom are treated on equal footing.
In \cite{bpp}, it was shown that,
  within relativistic string theory, these two
  methods lead to different spectra\footnote{
 ``The heuristic formulation'' of a conformally
 invariant theory \cite{poly}
 leads to conformal anomalies, the Virasoro algebra and
 tachyons (particles for which the masses squared are
 negative), at the same time eliminating extra degrees of
 freedom with a negative probability at the level of the classical
 theory \cite{bc} leads to the Born-Infeld model featuring a positive
 energy spectrum free from tachyons.}.

  Experience gained in applying the fundamental-quantization method
   \cite{sch2,pol} to string models  \cite{bpp,bc} and to non-Abelian theories
    \cite{mpl} shows that, upon a
spontaneous breakdown of the gauge symmetry of the theory being
considered, there arise Goldstone modes that are associated with
this symmetry breaking and which cannot be removed by any gauge
transformations without significantly changing the physical
content of the theory, including the spectrum of its elementary
and collective excitations\footnote{In the non-Abelian theory of
strong interactions,
    the analogous Goldstone mode leads to an extra contribution
    to the $\eta^0$-meson mass, while averaging over the topological
    degeneracy of initial data may lead to zero
    probabilities of the production of color states of quarks and
    gluons \cite{mpl}.}.

 In \cite{ps1,039,plb}, a spontaneous breakdown of the
scale invariance of the theory being considered is formulated in
terms of conformal variables, the Goldstone mode corresponding to
this symmetry breaking being taken into account.

\newpage

\centerline {\it 2.3. Conformal Variables}

\vspace{5mm}

 The choice of {\it ``relative'' (r)} variables,
\be \label{ccr} {}^{(n)}F_{(r)}={}^{(n)}F \times
(w/\vh(x^0))^{n},~~~~~~~~ w_{(r)}(x^0,x^i)=\vh(x^0), \ee
 leaves the dilaton zero mode as a homogeneous variable
 $\vh(x^0)$ with a constant volume of three-dimensional hyperspace \cite{ppgc,plb},
 $V_{(r)}=\int d^3 x |\bar e_{(r)}|\equiv{\rm const},$
   in the reference frame specified by the embedding
  of a three-dimensional hyperspace into the four-dimensional
  manifold spanned by  $e_{\underline{0}0}=N$, $e_{\underline{i}j}=
\bar e_{\underline{i}j}$, $e_{\underline{i}0}=N_{\underline{i}}$,
where $N$ and $N_{\underline{i}}$ are referred to the lapse
function and the shift vector \cite{vlad}, respectively. In this
reference frame, an variable  $\vh$ plays the role of a cosmic
scale factor and an evolution parameter in the world space of the
field variables $[\vh|F]$, while the canonical momentum defined as
the derivative of the Lagrangian $L_{\rm tot}$ for the action
functional (\ref{GR:CI})
 with respect to the time derivative of the dilaton field $\partial_0 \vh
 $,
  \be\label{GR:CI2}%
P_\vh = \frac{\partial L_{\rm tot}[\vh|e_{(r)}]}{\partial (\partial_0\vh)}
=-2\partial_0 \vh \int d^3x \frac{|\bar{e}_{(r)}|}{N}
\equiv -2 V_{(r)}\frac{d\vh}{d\eta},
\ee
 is the localized energy of the Universe  \cite{bpp}; here,
$d\eta =N_0(x^0)dx^0$ is the invariant integral for the averaged
lapse function $ N^{-1}_0(x^0)=\int d^3x
|\bar{e}_{(r)}|N^{-1}/V_{(r)} $, while the bar over
$\bar{e}_{(r)}$, is, as we have seen above, denotes the spatial
components of the vierbein for the constant volume  $V_{(r)}=\int
d^3 x |\bar e_{(r)}|$.

In terms of the conformal variables in  (\ref{ccr}),  the problems
of the theory that are solved in terms of the absolute variables
in  (\ref{cca}) with the aid of inflation are solved within the
exact theory by means of the zero mode of the dilaton as the
evolution parameter. In particular, the evolution of $\vh$ with
respect to the time interval explains the problem of the horizon
as a consequence of simultaneously varying particle masses and
parameters of the system of fields over the entire space. The
averaging of the exact equation  $\delta S_{\rm tot}/\delta N=0$
of the theory for the lapse function $N$ in terms of the variables
in (\ref{ccr}) over a specific spatial volume housing measured
objects yields the equation of the evolution of the Universe
   \be\label{GR2}
   \vh'^2=\rho,
 \ee
 where $\rho = \int d^3x |e|[T_0^0-\vh^2(R^0_0-R/2)]$, and $T_0^0$ и
 $R^0_0$ being the components of
 the Einstein energy-momentum and the Ricci tensor, respectively.

We note that, in the exact theory specified by the action
functional in (\ref{GR:CI}), Eq. (\ref{GR2}) is the analog of the
Friedmann equation, which was derived in the approximation of
homogeneity in the general theory of relativity. Thus, the
approximation of a homogeneous universe coincides with the result
obtained by averaging the exact equation $\delta S_{\rm
tot}/\delta N=0$ over the volume. Solving Eq. (\ref{GR2}), we
arrive at an analog of the Friedman relation between the conformal
time and density, $ \eta(\vh_0,\vh_I)= \pm
\int\limits_{\vh_I}^{\vh_0} {d\vh}/{\sqrt{\rho}} $.

Within the Hamiltonian formalism, where the time derivatives of
the fields are replaced by the corresponding canonical momenta
$P_\vh=-2V_{(r)}\vh'$,  the equation of the Universe evolution
(\ref{GR2}), which describes the evolution of the Universe, has
the meaning of a Hamiltonian constraint,
$P_\vh^2/4V_{(r)}=V_{(r)}\rho$. The Hamiltonian  $V_{(r)}\rho$ as
a generator of the conformal time evolution of fields can be
represented as the sum of the Hamiltonian for a uniform scalar
field,
 $V_{(r)}\rho_Q$, and the Hamiltonian for local field variables,
  $H_{\rm field}$:
 $V_{(r)}\rho=V_{(r)}\rho_Q+H_{\rm field}$, where $\rho_Q$
 is the density of the uniform scalar field.

In terms of the conformal variables in   (\ref{ccr}), the Planck
mass as the ``absolute'' parameter of the equations of motion
becomes a ``random'' current value of the field evolution
parameter $\vh(x^0)$. Hence, in terms of the variables in
(\ref{ccr}) both the hypothesis of the Planck era and the problem
of describing evolution from the Planck era $10^{-43} \mbox{\rm
c}$ on the basis of the inflationary model of the Universe lose
physical meaning\footnote{ The variables in  (\ref{cca}) arise
from (\ref{ccr})
 upon the substitution  ${}^{(n)}F_{(r)}={}^{(n)}F_{(a)} (\vh/\vh_0)^{-n}$.
 This transformation converts the variable $\vh$
 with initial cosmological data
 $\vh(\eta=0)=\vh_I$, $H(\eta=0)=H_I$  into its current value
 $\vh(\eta=\eta_0)=\vh_0$, with the result that one of the
 ordinary (random) values of the variable
  $\vh$
 becomes, for the equations of motion, the absolute
parameter $\vh_0=\sqrt{3/(8\pi)}M_{\rm Pl}$,
 which is related to the Planck mass.
}.

Within the conformal variables  (\ref{ccr}), there arises the
problem of studying the quantum creation and evolution of a
relativistic universe in the limit of infinitely low masses
$(\vh(\eta) \to 0)$ and indefinitely high values of the Hubble
parameter $(H(\eta) \to \infty)$. The variables in  (\ref{cca})
and those in (\ref{ccr}) provide two different cosmologies and two
different formulations of the problem of studying the origin of
the Universe and matter.

\vspace{10mm}

\centerline{\it 2.4. Conformal Cosmology}

\vspace{5mm}

In the approximation of homogeneity, the conformal variables in
(\ref{ccr})  correspond to directly measured quantities of
observational cosmology. We recall that, in describing the cosmic
evolution of the energy of photons emitted by atoms in a cosmic
object, use is made of the conformal interval of photons
$(dx^i)^2={dt}^2/a^2(t)= d\eta^2$ propagating along the light
cone, $ds^2=0$,
 toward an observer.
The redshift of spectral lines, which is a directly measurable
quantity in observational cosmology, depends on the ``conformal''
time $\eta=\eta_0-r$  at the instant of photon emission by atoms
of cosmic objects that occur at the ``coordinate distance''
$r=\sqrt{(x^i)^2}$ from the Earth.
 In terms of the conformal coordinates, we find that the volume
 of the Universe does not increase,
 while all masses, including the Planck mass, are scaled by
 the cosmic factor $a(\eta)$:
\be \label{mass} m_{(r)}(\eta)=m_0a(\eta),~~~[M_{\rm Pl}\sqrt{3/
8\pi}]a(\eta)=\vh_0a(\eta)=\vh(\eta). \ee In terms of the
conformal time, which is associated with the observed time, the
square-root regime of the evolution of the Universe in the era of
primordial nucleosynthesis, \be \label{rigid} a(t)=
\widetilde{a}(\eta) =\sqrt{1+2H_0(\eta-\eta_0)}=1- r H_0 +O(r^2)
  \ee
means that, in the era of chemical evolution, the Universe was
filled with a free uniform scalar field (see Eq. (\ref{cuf11})
below) rather than with radiation.  The evolution according to Eq.
(\ref{rigid}) is prescribed by a rigid equation of state such that
pressure coincides with the energy density.

 In \cite{039}, it was shown that, in terms of the conformal variables,
data on the dependence of the redshift on the distance to
supernovae \cite{ps} and data on nucleosynthesis correspond to the
same rigid equation of state associated with Eq. (\ref{rigid}).

The identification of the conformal variables in  (\ref{ccr}) with
observable quantities leads to a different picture of the
evolution of the Universe~\cite{ppgc,039,plb} in relation to the
analogous identification of the variables in (\ref{cca}) as is
done in conventional cosmology. The temperature history of a hot
universe as rewritten in terms of the conformal variables in
(\ref{ccr}) appears as the evolution of elementary particle masses
in a cold universe with a constant temperature of cosmic microwave
background radiation. That the cosmic microwave background
radiation temperature $ T_{\rm CMBR}$ is independent of the
redshift $z$ is, at first glance, in glaring contradiction with
the observation
 \cite{sria} that
$6.0~{\rm K} < T _ {\rm CMBR} (z=2.3371) < 14 ~ {\rm K} $. In this
observation, the temperature was deduced from the relative
population of various energy levels (their energies being denoted
by $E_i$), which follows from Boltzmann statistics. However, the
argument of the Boltzmann factors, which is equal to the ratio of
the temperature to the mass, features the same dependence on the
factor $z$  in a cold universe as well \cite{039}. Therefore, this
ratio can be interpreted as the $z$ dependence of energy levels
(that is, mass) at a constant temperature. The abundances of
chemical elements are determined primarily by Boltzmann factors as
well, which are dependent on functions of the mass-to-temperature
ratio, which are invariant under conformal transformations
\cite{three}.

\vspace{10mm}

\centerline {\it 2.5. Initial Data of Quantum Cosmology}

\vspace{5mm}

 As a rule, quantum cosmology is defined as the homogeneous
 approximation of the metric,
\be\label{conf1}
ds^2_{(r)}=[(d\eta)^2-(dx^i)^2],~~~~~d\eta=N_0(x^0)dx^0, \ee with
the shift function  $N_0(x^0)$ inheriting the symmetry group of
the general theory of relativity in the form of invariance under
reparametrizations of the coordinate time, $x^0 \to
\widetilde{x}^0=\widetilde{x}^0(x^0)$.
 Cosmological models featuring this symmetry group, which were
 first described at a mathematically rigorous level
 by DeWitt, Wheeler, and Misner
 \cite{dw,M} in the late 1960s, do not differ in any respect from
 the relativistic mechanics of a particle in
 the special theory of relativity.
There is a direct correspondence between the Minkowski space of
variables in the special theory of relativity and the space of
field variables in the theory being considered, where the dilaton
field
 $\vh$
 plays the role of the time-like variable of Minkowski space.

 In the particular case where the uniform scalar field
  $Q(\eta)$ 
is dominant, we arrive at a simple cosmological model of the
Universe  \cite{ppgc}; that is, \be\label{cuf} S_{\rm
univ}=V_{(r)} \int \frac{dx^0}{N_0}\left[
-\left(\frac{d\vh}{dx^0}\right)^2+\vh^2\left(\frac{dQ}{dx^0}\right)^2\right]=
\ee
$$=\int \left\{ -P_{\vh}\frac{d\vh}{dx^0}+P_Q\frac{dQ}{dx^0}+
{N_0}{V_{(r)}} \left[(P_\vh/2V_{(r)})^2-(P_Q/2\vh
V_{(r)})^2\right]\right\},
$$
where $P_Q=2V_{(r)}\vh^2 Q'$,  $P_\vh=2V_{(r)} \vh'$ are canonical
momenta.
 A variation of the action functional with respect to the shift function
  $N(x^0)$ leads to a constraint equation for these two momenta,
   \be\label{cuf1} P^2_\vh-
P^2_Q/\vh^2 =0~~~\Rightarrow~~~P_\vh=\pm P_Q/\vh, \ee  its
solution being \be\label{cuf11} P_Q=2V_{(r)}H_I\vh_I^2=\mbox{\rm
const},~~~~~\vh^2=\vh_I^2(1+2H_I\eta), \ee where $\vh_I^2H_I=P_Q/2
V_{(r)}$ is an integral of the motion. As was shown in \cite{039},
the resulting evolution law (\ref{cuf11}) for the scale factor is
compatible with the evolution of supernovae in terms of conformal
variables \cite{ps}.

Upon the quantization of the theory specified by the action
functional (\ref{cuf}), the Wheeler-DeWitt equation for the wave
function $\Psi$,
$$
\left[\hat P^2_\vh- \hat P^2_Q/\vh^2\right]\Psi =0
$$
arises as the direct analogy of the Klein--Gordon equation in the
quantum general theory of relativity.
 Like that solution to the Klein-Gordon equation for
a relativistic particle which describes the creation and
annihilation of positive-energy particles, the solution
$$
\Psi=A_{{\rm P}_\vh \geq 0}^+\Psi^+[P_Q|\vh]
e^{\{ iP_Q(Q-Q_I)\}}\theta(\vh-\vh_I)+
$$
$$
A_{{\rm P}_\vh \leq 0}^- \Psi^-[P_Q|\vh]e^{\{- iP_Q(Q-Q_I)\}}\theta(-\vh)
$$
to the Wheeler-DeWitt equation depends on the initial data  $Q_I$
and $\vh_I$.
 In order to get rid of negative energies and to create a stable quantum system,
  a causal quantization is postulated in quantum field theory
   in such a way that positive-energy excitations move
 in the forward direction along the time axis,
  while negative-energy excitations move
  in the backward direction with respect to time.
  The analogous interpretation of the coefficient
   $A^+$ as the creation operator for the Universe, and the coefficient
  $A^-$ as the annihilation operator for the anti-Universe solves
  the problem of a cosmic singularity of the Universe at positive ``energy'',
  since,
  for positive energies, the wave function does not involve the
  singularity point $\vh=0$; this singularity appears in the
  negative-``energy'' wave function, which is treated
  as the amplitude of the probability for the annihilation
  of the anti-Universe.

 The quantum general theory of relativity loses the
geometric interval of time and information about how the metric
depends on the time interval-in particular, it loses the Hubble
law describing the dependence of the scale on the time interval
[see Eqs. (\ref{cuf11})].
 In \cite{pp}, it was proposed to make a canonical transformation
(known as the Levi-Civita transformation) from the original
variables to a new world space of variables.
 In terms of these variables, the cosmic scale
 (dilaton) becomes a geometric interval of time with cosmic
 initial data $\vh(\eta=0)=\vh_I,~H(\eta=0)=H_I$  in  (\ref{cuf11}),
which are random values of variables fitted to experimental data.

In this case, the conformal variables in  (\ref{ccr}) naturally
lead to the concept of particles  \cite{ps1}, which has been used
and is being presently used in almost all of the studies devoted
to the cosmological creation of particles  \cite{grib80}.

\vspace{10mm}

\centerline {\it 2.6. Definition of a Particle in Quantum Field
Theory}

\vspace{5mm}

In quantum field theory, the concept of a particle can be
associated only with those field variables that are characterized
by a positive probability and a positive energy. Negative energies
are removed by causal quantization, according to which the
creation operator at a negative energy is replaced by the
annihilation operator at the respective positive energy. All of
the variables that are characterized by a negative probability can
be removed according to the scheme of fundamental operator
quantization \cite{sch2}. The results obtained by applying the
operator-quantization procedure to massive vector fields in the
case of the conformal flat metric (\ref{conf1}) are given in
\cite{ps1,ppgc,hp}.

In order to determine the evolution law for all fields  $\bf \rm
v$,
 it is convenient to use the Hamiltonian
 form of the action functional for their Fourier components
${\bf \rm v}_k^{I}=\int\limits_{} d^3xe^{\imath\bf k\cdot x}{\bf
\rm v}^{I}({\bf x})$; that is,
\begin{eqnarray}
\label{grad}
S_{\rm tot}=\int\limits_{x^0_1 }^{x^0_2 }dx^0
\left\{ \sum\limits_{k}
 \left[{\bf p}_{k}^{\bot}\partial_0{\bf \rm v}_{k}^{\bot}
 + {\bf p}_{k}^{||}\partial_0{\bf \rm v}_{k}^{||}\right]
-  P_{a} \partial_0a +{N}_{0}\left[\frac{P_{a}^2}{4V_{(r)}\vh_0^2}-
{V_{(r)}\rho_{\rm tot}}
\right]\right\},
\end{eqnarray}
where ${\bf p}_{k}^{\bot},{\bf p}_{k}^{||}$ are the canonical
momenta for, respectively, the transverse and the longitudinal
component of vector bosons and $\rho_{\rm tot}$ is the sum of the
conformal densities of the scalar field obeying the rigid equation
of state and the vector field,
\begin{eqnarray}
\label{totgrad}
\rho_{\rm tot}(a)&=&\frac{\vh_0^2 H_0^2}{a^2}+\rho_{\rm v}(a),~~~~\\
\rho_{\rm v}(a)  &=&V_{(r)}^{-1}(H^{\bot} + H^{||}),
\label{totgrad1}
\end{eqnarray}
$H^{\bot}$ и $H^{{||}}$ being the Hamiltonians for a free
field,\footnote{In quantum field theory, observables that are
constructed from the above field variables form the Poincare
algebra \cite{hp,sch2,mpl}. Therefore, such a formulation, which
depends on the reference frame used, does not contradict the
general theory of irreducible and unitary transformations of the
relativistic group
  \cite{Schweber}.
}
\begin{eqnarray}
&H^{\bot} = \sum\limits_{k} \frac{1}{2}\left[{\bf p}_k^{\bot}{}^2 +
\omega^2 {\bf \rm v}_k^{\bot}{}^2\right]~,
\nonumber\\ [-8mm]&\label{grad1}
\\&\nonumber
H^{||} = \sum\limits_{k} \frac{1}{2}\left
[\left(\frac{\omega(a,k)}{M_{\rm v} a}\right)^2{\bf
p}_{k}^{||}{}^2 + (M_{\rm v} a)^2 {\bf \rm v}_{k}^{||}{}^2
\right].
\end{eqnarray}
Here, the dispersion relation has the form  $\omega(a,k) =
\sqrt{{\bf k^2} + (M_{\rm v} a)^2}$; for the sake of brevity, we
have also introduced the notation ${\bf p}_{k}^{||}{}^2\equiv {\bf
p}_{k}^{||}{}\cdot{\bf p}_{-k}^{||}{} $.

Within the reparametrization-invariant models specified by action
functionals of the type in (\ref{grad}) with the Hamiltonians in
(\ref{grad1}), the concepts of an observable particle and of
cosmological particle creation were defined in \cite{ps1}. We will
illustrate these definitions by considering the example of an
oscillator with a variable energy. Specifically, we take its
Lagrangian in the form
 \be \label{la1}
 {\cal L}=p_{\rm v}\partial_0 {\rm v}-N_0\frac{1}{2}[p^2_{\rm v}+\omega^2{\rm v}^2-\omega]+\rho_0(N_0-1)~.
 \ee
The quantity  $H_{\rm v}=[p^2_{\rm v}+\omega^2{\rm v}^2]/2$ has
the meaning of a ``conformal Hamiltonian'' as a generator of the
evolution of the fields
 $v$ and $p_v$
with respect to the conformal-time interval  $d\eta=N_0dx^0$,
where the shift function $N_0$ plays the role of a Lagrange
multiplier.  The equation for  $N_0$  introduces the density
$\rho_0=H_{\rm v}-\omega/2$ in accordance with its definition
adopted in the general theory of relativity. In quantum field
theory \cite{grib,ps1}, the diagonalization of precisely the
conformal Hamiltonian \be \label{2la2} H_{\rm
v}=\frac{1}{2}[p^2_{\rm v}+\omega^2{\rm v}^2]=\omega\left[\hat
N_{\rm part}+\frac{1}{2}\right]
 \ee
specifies   both   the   single-particle   energy
$\omega=\sqrt{{\bf k}^2+(M_va(\eta))^2}$ and the particle-number
operator \be \label{la2} \hat N_{\rm part}
=\frac{1}{2\omega}[p^2_{\rm v}+\omega^2{\rm v}^2]-\frac{1}{2}
 \ee
with the aid of the transition to the symmetric variables $p$ and
$q$ defined as \be \label{la3}
 p_{\rm v}= \sqrt{\omega}p=i\sqrt{\frac{\omega}{2}}(a^+-a),
 ~~~~~~~~{\rm v}=\sqrt{\frac{1}{\omega}}q=\sqrt{\frac{1}{2\omega}}(a^++a).
 \ee
In terms of the symmetric variables $p,q$ the particle-number
operator takes form  \be \label{la4}
 \hat N_{\rm part}= \frac{1}{2}[p^2+q^2]-\frac{1}{2}=a^+a.
 \ee
Upon going over to these variables in the Lagrangian in
(\ref{la1}), we arrive at  \be \label{la5}
 {\cal L}=p\partial_0q-pq\partial_0 \Delta^{\bot}-N_0\omega[\hat N_{\rm part}+1/2]~,
 \ee
where $\partial_0 \Delta^{\bot}=\partial_0\omega/2\omega$ and
where there appears sources of cosmic particle creation in the
form $pq=i[(a^+)^2-a^2]/2$. Here, we give a derivation of these
sources for transverse fields, whereas, for longitudinal fields
[see Eq.(\ref{grad1})], the analogous diagonalization of the
Hamiltonian leads to the factor $\partial_0
\Delta^{||}=\partial_0\vh/\vh-\partial_0\omega/2\omega$.

In order to diagonalize the equations of motion in terms of the
mentioned new variables, it is necessary to apply, to the phase
space, the rotation transformation  \be \label{la6}
 p=p_\theta \cos\theta + q_\theta \sin\theta,~~~~~
 q=q_\theta \cos\theta - p_\theta \sin\theta
 \ee
and the squeezing phase space transformation  \be \label{la7}
 p_\theta =\pi e^{-r},~~~~~~~~q_\theta =\xi e^{+r}.
 \ee
As a result, the Lagrangian in (\ref{la5}) assumes the form  \be
\label{la8}
 {\cal L}=\pi\partial_0\xi+\pi\xi[\partial_0r-\partial_0 \Delta\cos2\theta]+
 \ee
 $$ +\frac{\pi^2}{2}e^{-2r}[\partial_0\theta-N_0\omega-
\partial_0 \Delta\sin2\theta ]+
\frac{\xi^2}{2}e^{2r}[\partial_0\theta-N_0\omega+
\partial_0 \Delta\sin2\theta ].
$$
The equations of motion that are obtained from this Lagrangian,
\be \label{la9}
 \xi'+\xi[r'- \Delta'\cos2\theta]+
  {\pi}e^{-2r}[\partial_0\theta-N_0\omega-
\partial_0 \Delta\sin2\theta ]=0,
\ee
\be \label{la10}
 \pi'-\pi[r'- \Delta'\cos2\theta]
-{\xi}{2}e^{2r}[\partial_0\theta-N_0\omega+
\partial_0 \Delta\sin2\theta ]=0,
 \ee
take a diagonal form, \be \label{la11}
 \xi'+\omega_b\pi=0,~~~~~~~~~-\pi'+\omega_b\xi=0,
 \ee
if $\omega_b=e^{-2r}[\omega-\theta'-\Delta'\sin2\theta]$ and if
the rotation parameter  $\theta$ and the squeezing parameter $r$
satisfy the equations\footnote{These equations for transverse and
longitudinal bosons coincide completely with the equations for the
coefficients of the Bogolyubov transformation $b=\alpha a+\beta
a^+$,~~ $\alpha'-i\omega\alpha=\Delta'\beta$,
 derived by using the Wentzel-Kramers-Brillouin method in \cite{grib80},
{see Eqs. (9.68) and (9.69) in [7] on page 185 in the Russian
edition of this monograph}, where it is necessary to make the
change of variables specified by the equations $
\Delta'=\frac{\omega^{(1)}}{2},~ \alpha^*= \exp[i\theta-i\int
d\eta \omega]{\rm ch} r,~ \beta= \exp[-i\theta+i\int d\eta
\omega]{\rm sh} r. $} \be \label{la12} [\theta'-\omega]{\rm sh}2r
=-\Delta'\sin2\theta {\rm ch}2r,~~~~~~r'=\ \Delta'\cos2\theta .
 \ee
By solving these equations, we can find the time dependence
 of the number of particles produced in cosmic evolution (\ref{la4})
 \be \label{la13} \hat
N_{\rm part}=\frac{{\rm ch}2r-1}{2}+{\rm ch}2r\hat N_{\rm q-part}+
{\rm sh}2r\frac{\pi^2-\xi^2}{2},
 \ee
where $\hat N_{\rm q-part}=[\pi^2+\xi^2-1]/{2}=b^+b$ is the number
of quasiparticles defined as variables that diagonalize the
equation of motion. Since the equation of motion is diagonal, the
number of quasiparticles is an integral of the motion, that is, a
quantum number that characterizes the quantum state of the
Universe.  One of these states is the physical vacuum state
$|0\rangle_{\rm sq}$ of quasiparticles (that is, the squeezed
vacuum, which is labelled with the subscript  ``sq'' in order to
distinguish it from the vacuum of ordinary particles),
\be\label{vacuum} b_{\varsigma}|0\rangle_{\rm sq} =
0~~~~~(b=\frac{1}{\sqrt{2}}[\xi+i\pi]). \ee In the squeezed-vacuum
state, the number of quasiparticles is equal to zero \be
\label{la14}
{}_{\rm sq}\langle 0|\hat N_{\rm q-part}|0\rangle_{\rm sq}=0.
 \ee
In this case, the expectation value of the particle-number
operator (\ref{la13}) in the squeezed-vacuum state is \be
\label{la15}
{}_{\rm sq}\langle 0|\hat N_{\rm part}|0\rangle_{\rm
sq}=\frac{{\rm ch}(2r(\eta))-1}{2} ={\rm sh}^2r(\eta).
 \ee
The time dependence of this quantity is found by solving the
Bogolyubov equation  (\ref{la12}). The origin of the Universe is
defined as the conformal-time instant $\eta=0$,
 at which the number of particles and the number of
 quasiparticles are both equal to zero.
The resulting set of Eqs. (\ref{la12}) becomes closed upon
specifying the equation of state and initial data for the number
of particles. In just the same way, the number of particles
characterized by an arbitrary set of quantum numbers  $\varsigma$,
\be {\cal N}_{\varsigma}(\eta) = {}_{\rm sq} \langle 0|\hat
N_{\varsigma}|0\rangle_{\rm sq} = {\rm sh}^2
r_{\varsigma}(\eta),\nonumber \ee and produced from the
``squeezed'' vacuum by the time instant $\eta$ can be determined
by solving an equation of the type in (\ref{la12}).

Thus, just the conformal quantities of the theory, such as the
energy  $\omega_k=\sqrt{k^2+M^2_va^2}$, the number particles $\hat
N_{\rm part}$, the conformal density $$\rho_{\rm {\rm v}} = \sum_k
{}_{\rm sq}\langle 0|\hat N_{k ~\rm part}|0\rangle_{\rm
sq}\omega_k/V_{(r)}$$ that are associated with observables,
  in just the same way as the conformal time in
  observational cosmology is associated with
  the observed time \cite{039}.

\vspace{20mm}

\centerline{\it 3. PHYSICAL IMPLICATIONS }


\centerline {\it 3.1. Calculation of the Distribution Function}

\vspace{5mm}

Let us consider the example where the above set of equations is
solved for the evolution law  (\ref{rigid}) in the case of the
rigid equation of state,
 $$a(\eta)=a_I\sqrt{1+2H_I\eta}~~~~~(a^2_IH_I=H_0),$$
where $a_I=a(0)$ and $H_I$ are initial data at the
matter-production instant.

We introduce the dimensionless variables of time  $\tau$ and
momentum $x$ and the coefficient
 $\gamma_{\rm v}$ according to the formulas
\be\label{g}
 \tau=2\eta H_I=\eta/\eta_I,~~~~~~~~ x=\frac{q}{M_{I}},
 ~~~~~~~~\gamma_{\rm v}=\frac{M_{I}}{H_I},
\ee where $M_{I}=M_{\rm v}(\eta=0)$  are initial data for the
mass.  In terms of these variables, the single-particle energy has
the form
 $\omega_{\rm v}=H_I\gamma_{\rm v}\sqrt{1+\tau+x^2}$.

The Bogolyubov equations~(\ref{la12}) can be represented as
$$
\left[\frac{\gamma_{\rm v}}{2}\sqrt{(1+\tau)+x^2} -
\frac{d\theta^{||}_{\rm v}}{d\tau}\right] \tanh(2r^{||}_{\rm v}) =
-\left[\frac{1}{2(1+\tau)}-\frac{1}{4\left[ (1+\tau)+x^2\right]}\right]
\sin(2\theta^{||}_{\rm v}),
$$
$$
\frac{d}{d\tau}r^{||}_{\rm v} =
\left[\frac{1}{2(1+\tau)}-\frac{1}{4\left[ (1+\tau)+x^2\right]}\right]
\cos(2\theta^{||}_{\rm v}),
$$
$$
\left[\frac{\gamma_{\rm v}}{2}\sqrt{(1+\tau)+x^2} -
\frac{d}{d\tau}\theta^{\bot}_{\rm v}\right]  \tanh(2r^{\bot}_{\rm v}) =
-\left[\frac{1}{4\left[ (1+\tau)+x^2\right]}\right]
\sin(2\theta^{\bot}_{\rm v}),
$$
\be\label{1g} \frac{d}{d\tau}r^{\bot}_{\rm v} =
\left[\frac{1}{4\left[ (1+\tau)+x^2\right]}\right]
\cos(2\theta^{\bot}_{\rm v}). \ee We solved these equations
numerically at positive values of the momentum $x=q/M_I$,
considering that, for $\tau\to+0$, the asymptotic behavior of the
solutions is given by $r(\tau)\to{\rm const}\cdot\tau$ and
$\theta(\tau)=O(\tau)$.
 The distributions of
longitudinal ${\cal N}^{||}(x,\tau)$ and transverse ${\cal
N}^{\bot}(x,\tau)$ vector bosons are given in the Figure 1. for
the initial data $H_I=M_I~(\gamma_{\rm v}=1$).

From the Figure 1, it can be seen that, for $x>1$, the
longitudinal component of the boson distribution is everywhere
much greater than than the transverse component, this
demonstrating a more copious cosmological creation of longitudinal
bosons in relation to transverse bosons. A slow decrease in the
longitudinal component as a function of momentum leads to a
divergence of the integral for the density of product
particles~\cite{par}: \be\label{nb}
 n_{\rm v}(\eta)=\frac{1}{2\pi^2}
\int\limits_{0 }^{\infty }
dq q^2
\left[ {\cal N}^{||}(q,\eta) + 2{\cal N}^{\bot}(q,\eta)\right]\to\infty.
\ee
\begin{figure}[t]
  \centering
  \includegraphics[width=0.45\textwidth,height=0.50\textwidth,angle=-90]{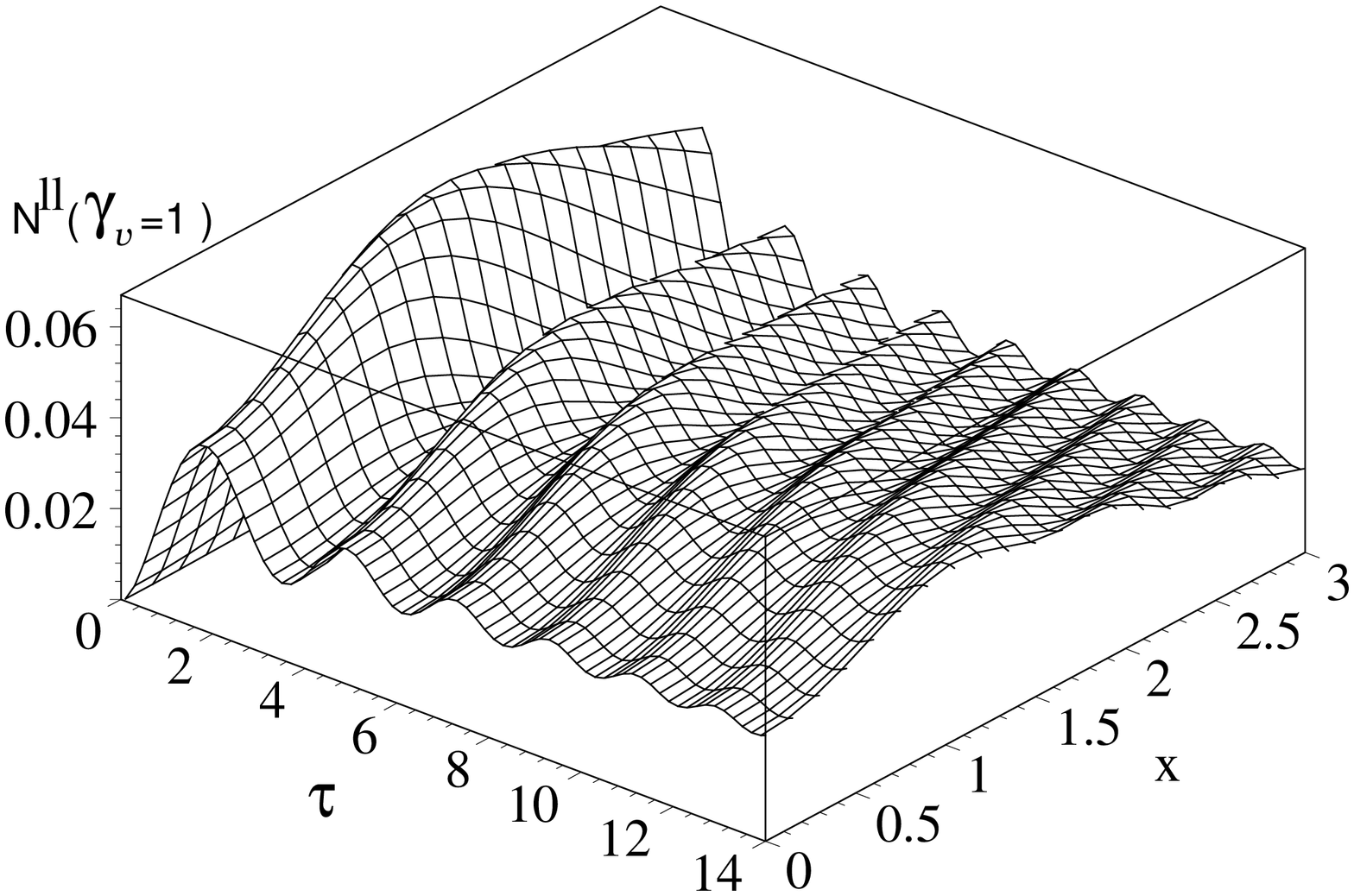}\hspace{-5mm}
  \includegraphics[width=0.45\textwidth,height=0.50\textwidth,angle=-90]{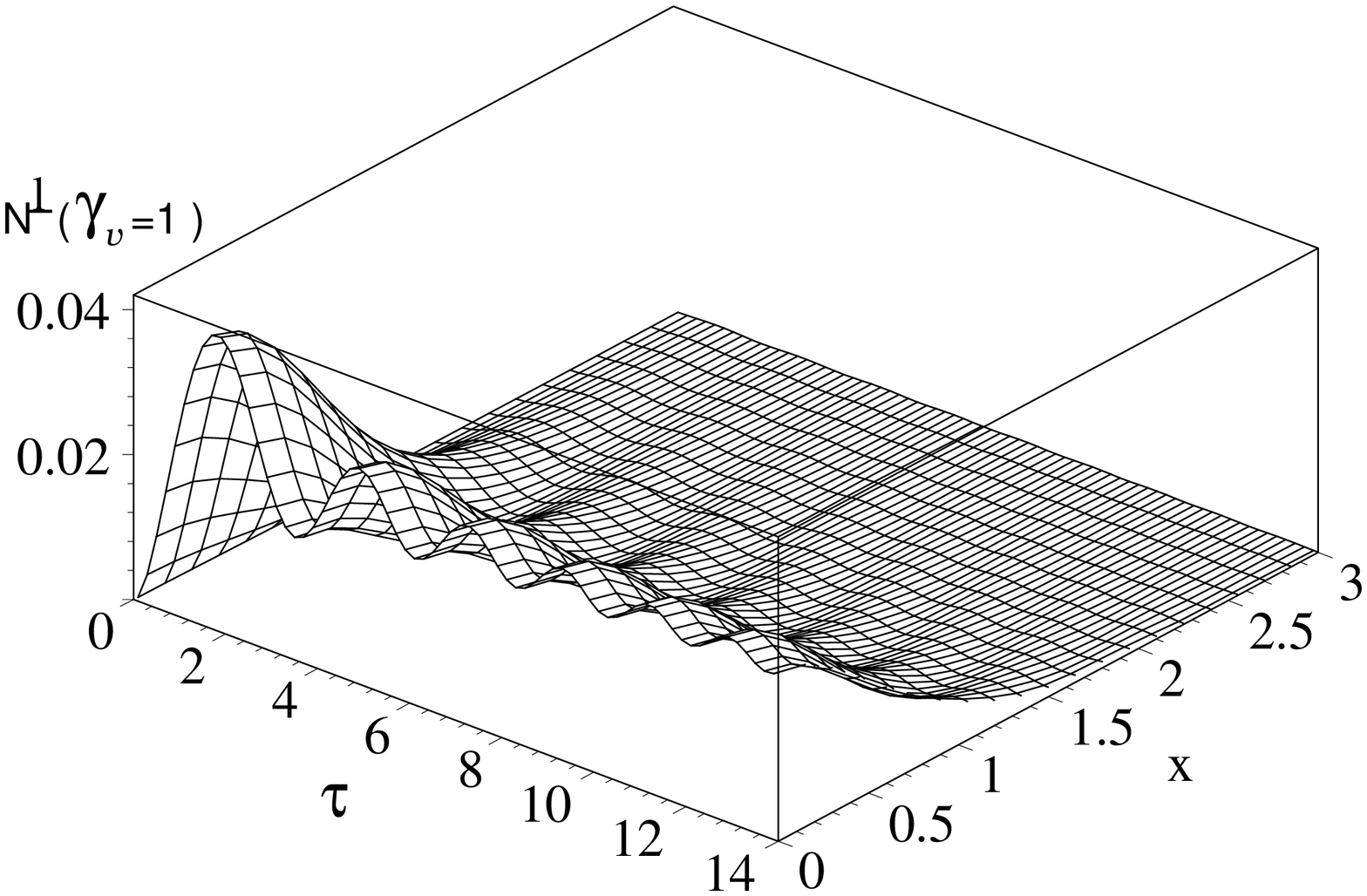}
  \caption{The dependence of the longitudinal $N^\|$ and transversal $N^\bot$
components of the boson distribution function on dimensionless
time  $\tau=2\eta H_I$ and dimensionless momentum~$x = q / M_I$ at
the initial data value~~$M_I=H_I$ ($\gamma_{\rm v} = 1$).}
\end{figure}
\vspace{1cm}

\centerline{\it 3.2. Thermalization of Bosons }

\vspace{5mm}

The divergence of the integral (\ref{nb}) stems from idealizing
the problem of the production of a pair of particles in a finite
volume for a system where there are simultaneous interactions
associated with the removal of fields having a negative
probability and where identical particles affect one another
(so-called exchange effects). In this case, it is well
known~\cite{ll}, that one deals with the production not a pair but
a set of Bose -- particles, which acquires, owing to the
aforementioned interactions, the properties of a statistical
system.   As a model of such a statistical system, we consider
here a degenerate Bose-Einstein gas, whose distribution function
has the form  (we use the system of units where the Boltzmann
constant is
 $k_{\rm B}=1$) \be \label{bose} {\cal F}\left(T_{\rm v},q,M_{\rm v}(\eta),\eta\right)=
 \left\{\exp\left[\frac{\omega_{\rm v}(\eta)- M_{\rm v}(\eta)}{ T_{\rm v}}\right]
 -1\right\}^{-1}, \ee
 where $T_{\rm v}$ is the boson temperature.
 We set apart the problem of theoretically validating
 such a statistical system and its thermodynamic exchange,
 only assuming fulfillment of specific conditions ensuring
 its existence. In particular, we can introduce the notion of the temperature
  $T_{\rm v}$ only in an equilibrium system.
 A thermal equilibrium is thought to be stable if the time within which the
 vector-boson temperature
 $T_{\rm v}$ is established, that is, the relaxation time \cite{ber} \be \label{rel01}
 \eta_{\mbox{\small  rel}} =
 \left[{n(T_{\rm v})\sigma_{\mbox{\small scat}}}\right]^{-1}
 \ee (expressed in terms of their density  $n(T_{\rm v})$
 and the scattering cross section
 $\sigma_{\mbox{\small  scat.}} \sim 1/ M_{I}^2$), does not exceed the
 time of vector-boson-density formation owing to cosmological creation,
 the latter time being controlled by the primordial Hubble parameter
  $\eta_{\rm v}=1/ H_I$. From formula (\ref{rel01}) it follows,
  that the particle-number density is proportional to the product of
  the Hubble parameter and the mass squared, that is
  an integral of the motion in the present example:
  \be \label{bose11}
 n(T_{\rm v})=n(T_{\rm v},\eta_{\rm v})\simeq
  C_H H_IM_I^2, 
\ee where $C_H$ is a constant. The expression for the density $
n(T_{\rm v},\eta)$ in Eq. (\ref{bose11}) assumes the form
\be\label{n1}
 n_{{\rm v}}(T_{\rm v},\eta)=\frac{1}{2\pi^2}
\int\limits_{0 }^{\infty } dq q^2{\cal F}\left(T_{\rm
v},q,M(\eta),\eta\right) \left[ {\cal N}^{||}(q,\eta) + 2{\cal
N}^{\bot}(q,\eta)\right]~. \ee Here, the probability of the
production of a longitudinal and a transverse boson with a
specific momentum in an ensemble featuring exchange interaction is
given (in accordance with the multiplication law for
probabilities) by the product of two probabilities, the
probability of their cosmological creation,  ${\cal N}^{||,\bot}$
and the probability of
 a single-particle state of vector bosons obeying
 the Bose-Einstein distribution~(\ref{bose}).

A dominant contribution to the integral (\ref{n1}) from the region
of high momenta (in the above idealized analysis disregarding the
Boltzmann factor, this resulted in a divergence)
 implies the relativistic temperature dependence of the density,
 \be \label{nc1}
 n(T_{\rm v},\eta_{\rm v}) = C_T T_{\rm v}^3,
 \ee
where $C_T$ is a coefficient. A numerical calculation of the
integral (\ref{n1}) for the values  $T_{\rm v}=M_I=H_I$, which
follow from the assumption about the choice of initial data
($C_T=C_H$), reveals that this integral (\ref{n1}) is weakly
dependent on time in the region  $\eta \geq \eta_{\rm v}=H_I^{-1}$
and, for the constant $C_T$, yields the value \be \label{nc} C_T =
\frac{ n_{\rm v}}{ T_{\rm v}^3} = \frac{1}{2\pi^2} \left\{
[1,877]^{||}+2 [0,277]^{\bot}=2,431 \right\}~, \ee where the
contributions of longitudinal and transverse bosons are labeled
with the superscripts  $(||,~ \bot)$, respectively.

 On the other hand, the lifetime $\eta_L$ of
 product bosons in the early Universe in dimensionless units
$\tau_L=\eta_L/\eta_I$, where $\eta_I=(2H_I)^{-1}$, can be
estimated by using the equation of state
$a^2(\eta)=a_I^2(1+\tau_L)$ and the $W$-boson lifetime within the
Standard Model. Specifically, we have \be \label{life} 1+\tau_L=
\frac{2H_I\sin^2 \theta_{(W)}}{\alpha_{\rm QED} M_W(\eta_L)}=
\frac{2\sin^2 \theta_{(W)}}{\alpha_{\rm QED}\gamma_{\rm
v}\sqrt{1+\tau_L}}, \ee where $\theta_{(W)}$ is the Weinberg
angle,  $\alpha_{\rm QED}=1/137$ is the fine-structure constant,
and $\gamma_{\rm v}=M_{I}/ H_I\geq 1$.

From the solution to Eq.~(\ref{life}), \be \label{lifes} \tau_L+1=
\left(\frac{2\sin^2\theta_{(W)}}{\gamma_{\rm v}\alpha_{\rm
QED}}\right)^{2/3} \simeq \frac{16}{\gamma_{\rm v}^{2/3}}~ \ee it
follows that, at $\gamma_{\rm v}=1$, the lifetime of product
bosons is an order of magnitude longer than the Universe
relaxation time: \be \label{lv} \tau_L
=\frac{\eta_L}{\eta_I}\simeq \frac{16}{\gamma_{\rm v}^{2/3}}-1=15.
\ee

Therefore, we can introduce the notion of the vector-boson
temperature $T_{\rm v}$, which is inherited by the final vector
boson decay products (photons). According to currently prevalent
concepts, these photons form cosmic microwave background radiation
in the Universe. Indeed, suppose that one photon comes from the
annihilation of the products of $W^\pm$-boson decay and that the
other comes from  $Z$-bosons. In view of the fact that the volume
of the Universe is constant within the evolution model being
considered, it is then natural to expect that the photon density
coincides with the boson density \cite{ppgc} \be \label{1nce}
n_\gamma={T_\gamma^3}\frac{1}{\pi^2} \left\{ 2.404 \right\} \simeq
n_{\rm v}. \ee

On the basis of  (\ref{bose11}), (\ref{nc1}), (\ref{nc}) and
(\ref{1nce}) we can estimate the temperature $T_{\gamma}$ of
cosmic microwave background radiation arising upon the
annihilation and decay of $W^+ $ and $Z$-bosons: \be \label{1nce1}
T_\gamma\simeq \left[\frac{ 2.431}{2.404 \cdot 2
}\right]^{1/3}T_{\rm v}=0.8 T_{\rm v}, \ee   taking into account
that the temperature of vector-bosons  $T_{\rm v}=
[H_IM^{2}_I]^{1/3}$ is an invariant quantity in the described
model. This invariant can be estimated at \be T_{\rm v}=
[H_IM^{2}_I]^{1/3}=[H_0 M_W^2]^{1/3}=2.73/0.8K=3.41K \ee which is
a value that is astonishingly close to the observed temperature of
cosmic microwave background radiation. In the present case, this
directly follows, as is seen from the above analysis of our
numerical calculations, from the dominance of longitudinal vector
bosons with high momenta and from the fact that the relaxation
time is equal to the inverse Hubble parameter. The inclusion of
physical processes, like the heating of photons owing to
electron-positron annihilation
  $e^+~e^-$  \cite{ee}
amounts to multiplying the photon temperature (\ref{1nce1}) by
$(11/4)^{1/3}=1.4$ therefore, we have \be \label{1nce11}
T_\gamma(e^+~e^-)\simeq (11/4)^{1/3}0.8 T_{\rm v}=2.77 ~K~. \ee We
note that, in other models  \cite{mar}, the fluctuations of the
product-particle density are related to primary fluctuations of
cosmic microwave background radiation   \cite{33}.

\vspace{1cm}

 \centerline{\it 3.3. Inverse Effect of Product
Particles on the Evolution of the Universe}

\vspace{0.5cm}

The equation of motion
 $\vh'^2(\eta)=\rho_{\rm tot}(\eta)$, with the
Hubble parameter defined as $H=\vh'/\vh$, means that, at any
instant of time, the energy density in the Universe is equal to
the so-called
 critical density; that is
$$
 \rho_{\rm tot}(\eta)= H^2(\eta)\vh^2(\eta)
\equiv \rho_{\rm cr}(\eta)~.
$$
The dominance of matter obeying the extremely rigid equation of
state implies the existence of an approximate integral of the
motion in the form
$$ H(\eta)\vh^2(\eta)=H_0\vh_0^2~.
$$

 On this basis, we can immediately find the ratio of the
 product-vector-boson energy, $ \rho_{\rm v}(\eta_I)\sim  T^4\sim
 H_I^4\sim M^4_{I}$, to the density of the Universe in the extremely
 rigid state,  $ \rho_{\rm tot}(\eta_I)=H_I^2\vh^2_{I}$,
 \be
 \label{F} \frac{\rho_{\rm v}(\eta_I)}{\rho_{\rm tot}(\eta_I)}=
 \frac{ M^2_{I}}{\vh_I^2} =\frac{M^2_{W}}{\vh_0^2}=y^2_{\rm
 v}=10^{-34}.
 \ee
 This value indicates that the inverse effect of product
particles on the evolution of the Universe is negligible.

{The primordial mesons before
 their decays polarize the Dirac fermion vacuum and give the
 baryon asymmetry frozen by the CP -- violation,
 so that $n_b/n_\gamma \sim X_{CP} \sim 10^{-9}$ and
 $\Omega_b \sim \alpha_{\rm qed}/\sin^2\theta_{(\rm W)}\sim
 0.03$.}

 \vspace{1cm}

\centerline{\label{3.4}\it 3.4. Baryon-antibaryon Asymmetry of
Matter in the Universe}

\vspace{0.5cm}

In each of the three generations of leptons
 (е,$\mu$,$\tau$) and color quarks, we have four fermion
 doublets-in all, there are $n_L=12$ of them. Each of 12 fermion
 doublets interacts with the triplet of non-Abelian fields
 $A^1=(W^{(-)}+W^{(+)})/\sqrt{2}$, $A^2=
 i(W^{(-)}-W^{(+)})/\sqrt{2}$, and $A^3=Z/\cos\theta_{(W)}$,
  the corresponding coupling constant being $g=e/\sin\theta_{(W)}.$

 It is well known that, because of a triangle anomaly, W- and
Z- boson interaction with lefthanded fermion doublets
 $\psi_L^{(i)}$, $i=1,2,...,n_L$, leads to
a nonconservation of the number of fermions of each type ${(i)}$
 ~\cite{bj,th},
 \bea \label{rub}
 \partial_\mu j^{(i)}_{L\mu}=\frac{1}{32\pi^2}
 {\rm Tr}\hat F_{\mu\nu}{}^*\!{\hat F_{\mu\nu}},
 \eea
 where $\hat
 F_{\mu\nu}=-iF^a_{\mu\nu}g_W\tau_a/2$ is the strength of the
 vector fields, $F^a_{\mu\nu}=
 \partial_\mu A_\nu^a-\partial_\nu
 A_\mu^a+g\epsilon^{abc}A_\mu^bA_\nu^c$.

 Taking the integral of the equality in (\ref{rub}) with respect to
 the four-dimensional variable $x$, we can find a relation between
 the change   $\Delta F^{(i)}=\int d^4x
 \partial_\mu j^{(i)}_{L\mu}$ the fermion number $ F^{(i)}=\int d^3x
 j_0^{(i)}$ and the Chern-Simons functional~\cite{ufn},
 $N_{CS}=\frac{1}{32\pi^2}\int d^4x {\rm Tr}\hat
 F_{\mu\nu}{}^*\!{\hat F_{\mu\nu}}$: \be \label{rub2}
 \Delta F^{(i)}= N_{CS} \not = 0, ~~~i=1,2,...,n_L.
 \ee
 The equality in (\ref{rub2}) is considered as a selection rule --
that is, the fermion number changes identically for all fermion
types:
 $N_{CS}=\Delta L^e=\Delta L^\mu=\Delta L^\tau=\Delta B/3$;
 at the same time, the change in the baryon charge $B$ and the change
 in the lepton charge  $L=L^e+L^\mu+L^\tau$ are related to each other in
such a way that $B-L$ is conserved, while
 $B+L$ is not invariant. Upon taking the sum of the equalities in
 (\ref{rub2}) over all doublets, we obtain $\Delta (B+ L)=12
 N_{CS}$.

 We can evaluate the expectation value of the Chern-Simons
 functional (\ref{rub2})  (in the lowest order of perturbation
 theory in the coupling constant) in the Bogolyubov vacuum
 $b|0>_{\rm sq}=0$. Specifically, we have
 \be
 N_{CS}=N_{\rm W}+N_{\rm Z}\equiv
 -\sum_{{\rm v}=W,Z}\int\limits_0^{\eta_{L_{\rm v}}} d\eta \int \frac{d^3 x}{32\pi^2} \;
 {}_{\rm sq}\langle 0|{\rm Tr}\hat F^{\rm v}_{\mu\nu}
 {}^*\!{\hat F^{\rm v}_{\mu\nu}}|0\rangle{}_{\rm sq} ,
 \ee
 where $\eta_{L_W}$ and  $\eta_{L_Z}$ are the W- and the Z-boson
 lifetime, and $N_{\rm W}$ and $N_{\rm Z}$
 are the contributions
 of primordial W and Z bosons, respectively.
 The integral over the conformal spacetime bounded
 by three-dimensional hypersurfaces $\eta=0$ and $\eta =\eta_L$
  is given by
 $$ 
 N_{\rm v} =\beta_{\rm v}\frac{V_0}{2}
 \int_{0}^{{\eta_{L_{\rm v}}}} d\eta \int\limits_{0 }^{\infty }dk
 |k|^3 R_{\rm v}(k,\eta)
 $$
 where ${\rm v}=W,Z$;
 \be\beta_W=\frac{4{\alpha}_{\rm
 QED}}{\sin^{2}\theta_{(W)}}, ~~\beta_Z=\frac{{\alpha}_{\rm
 QED}}{\sin^{2}\theta_{(W)}\cos^{2}\theta_{(W)}},\ee
 and the rotation parameter
 $$R_{\rm v}=-\sinh(2r)\sin(2\theta)
 $$
  is
 specified by relevant solutions to the
 Bogolyubov equations (\ref{1g}). Upon a numerical calculation
 of this integral, we can estimate the expectation value of the
 Chern-Simons functional in the state of primordial bosons.

 Primordial fluctuation of the baryon number
 \be\label{fluc}
 \frac{n_b(\eta)}{n_b(\eta_L)}=\frac{\int dk
 k^2\int\limits_{0}^{\eta}[4\cos^{2}\theta_{(W)}R_{\rm
 W}+R_{\rm Z}]}{\int dk k^2[
 4\cos^{2}\theta_{(W)}\int\limits_{0}^{\eta_w}d\eta R_{\rm
 W}+\int\limits_{0}^{\eta_z}d\eta R_{\rm Z}]}.
 \ee

 At the vector-boson-lifetime values of
 $\tau_{L_W}= 15$, $\tau_{L_Z}= 30$,
 this yields the following result at $n_\gamma\simeq n_{\rm v}$
\be
\frac{N_{CS}}{V_{(r)}}=\frac{( N_W+ N_Z)}{V_{(r)}}\\
 =\frac{{\alpha}_{\rm QED}}{\sin^{2}\theta_{(W)}}
  T^3
  \left(4\times 1.44+\frac{2.41}{\cos^{2}\theta_{(W)}}\right)
  =1.2~  n_{\gamma}.
\ee On this basis, the violation of the fermion-number densityin
the cosmological model being considered can be estimated as
\cite{039}
\begin{eqnarray}
\frac{\Delta F^{(i)}}{V_{(r)}}&=&\frac{N_{CS}}{V_{(r)}}
  =1.2  n_{\gamma},
\end{eqnarray}
 where $n_{\gamma}={ 2,402  \times T^3 }/{\pi^2}$ is the number
 density of photons forming cosmic microwave background radiation.

 According to Sakharov \cite{sufn} this violation of the
 fermion number is frozen by ${\rm CP}$ nonconservation, this
 leading to the baryon-number density
 \be\label{X} n_{\rm b}=
  X_{\rm CP}\frac{\Delta
  F^{(i)}}{V_{(r)}}\simeq X_{\rm CP}n_{\gamma}~.
  \ee
  where the factor $X_{\rm CP}$ is determined by the superweak
 interaction of $d$ and $s$ quarks,
 which
 is responsible for CP violation experimentally observed in
 $K$-meson decays \cite {o} (see Fig. 2).

 From the ratio of the number of baryons to the number of photons,
 one can deduce an estimate of the superweak-interaction coupling
 constant: $X_{\rm CP}\sim 10^{-9}$.
  Thus, the evolution of the Universe, primary
 vector bosons, and the aforementioned superweak interaction \cite{o}
 (it is responsible for CP violation and is characterized by a
 coupling-constant value of $X_{CP}\sim
10^{-9}$) lead to baryon-antibaryon
 asymmetry of the Universe, the respective baryon density being
 \be\label{data6} \rho_{\rm b}(\eta=\eta_{L})
 \simeq 10^{-9} \times 10^{-34}\rho_{\rm cr}(\eta=\eta_{L}).
 \ee
 In order to assess the
further evolution of the baryon density, one can take here the
W-boson lifetime for $\eta_{L}$.

 Upon the decay of the vector bosons in question,
their temperature is inherited by cosmic microwave background
radiation. The subsequent evolution of matter in a stationary cold
universe is an exact replica of the well-known scenario of a hot
universe \cite{three}, since this evolution is governed by
conformally invariant mass-to-temperature ratios $m/T$.

Formulas (\ref{life}), (\ref{F}), and (\ref{data6}) make it
possible to assess the ratio of the present-day values of the
baryon density and the density of the scalar field, which plays
the role of primordial conformal quintessence in the model being
considered. We have
  \be \Omega_{\rm
 b}(\eta_0)=\frac{\rho_{\rm b}(\eta_{0})}{\rho_{\rm cr}(\eta_{0})}=
 \left[\frac{\vh_0}{\vh_L}\right]^3=\left[\frac{\vh_0}{\vh_I}\right]^3
 \left[\frac{\vh_I}{\vh_L}\right]^3,
 \ee
 where we have considered that the baryon density increases in
 proportion to the mass and that the density of the primordial
 quintessence  decreases in inverse proportion to the mass
 squared. We recall that the ratio $[{\vh_0}/{\vh_I}]^3$
  is approximately equal
 to $10^{43}$ and that the ratio $[{\vh_I}/{\vh_L}]^3$ is
 determined by the boson
 lifetime in (\ref{lifes}) and by the equation of state
 $\vh(\eta)\sim \sqrt{\eta}$. On this basis, we can estimate
 $\Omega_{\rm b}(\eta_0)$ at
 \be\label{data7} \Omega_{\rm b}(\eta_0)
 =\left[\frac{\vh_0}{\vh_L}\right]^3 10^{-43}
 \sim 10^{43} \left[\frac{\eta_I}{\eta_L}\right]^{3/2}10^{-43}
 \sim \left[\frac{\alpha_{QED}}{\sin^2 \theta_{(W)}}\right] \sim
 0.03 ,
 \ee
 which is compatible with observational data \cite{fuk}.

 Thus, the general theory of relativity and the Standard Model,
 which are supplemented with a free scalar field   in a specific
 reference frame with the initial data $\vh_I=10^{4}$
 $H_I=2.7~{\rm K}$,
do not contradict the following scenario of the evolution of the
Universe within conformal cosmology \cite{039}:\\[1.5mm]
 $\eta \sim 10^{-12}s,$ {creation of vector bosons from a
 ``vacuum''};\\ [1.5mm] $10^{-12}s < \eta <
10^{-11}\div 10^{-10} s,$ {formation of baryon-antibaryon asymmetry;}\\
[1.5mm] $\eta \sim 10^{-10}s,$ {decay of vector bosons;}\\
[1.5mm] $10^{-10}c <\eta < 10^{11}s,$ { primordial chemical
evolution of matter;}\\ [1.5mm] $\eta \sim 10^{11}s,$
{recombination or separation of cosmic
microwave background radiation;}\\
[1.5mm] $\eta \sim  10^{15}s,$ {formation of galaxies;}\\
[1.5mm]  $\eta > 10^{17}s,$ { terrestrial experiments and
evolution of supernovae.}

\vspace{20mm}

\centerline {\bf 4. CONCLUSION}

\vspace{2mm}

Within the conformal formulation of the general theory of
relativity and the Standard Model, we have investigated conditions
under which the origin of matter can be explained by its
cosmological creation from a vacuum. We have presented some
arguments in support of the statement that the number of product
vector-boson pairs is sufficient for explaining the total amount
of observed matter and its content, provided that the Universe is
considered as a conventional physical object that is characterized
by a finite volume and a finite lifetime and which is described by
a conformally invariant version of the general theory of
relativity and the Standard Model featuring scale-invariant
equations where all masses, including the Planck mass, are
replaced by the dilaton variable and where the spatial volume is
replaced by a constant. In this case, the energy of the entire
Universe in the field space of events is described by analogy with
the description of the energy of a relativistic quantum particle
in Minkowski space: one of the variables (dilaton in the case
being considered) becomes an evolution parameter, while the
corresponding canonically conjugate momentum assumes the role of
energy \cite{pp,bpp}.
  This means that measured quantities are
  identified with conformal variables that are used
  in observational cosmology and in quantum field theory in calculating
  cosmological particle creation from a vacuum
 \cite{zel1,grib80,ps1,pp}.
Within the errors of observation, this identification of conformal
variables with observables is compatible with data on the chemical
evolution of matter and data on supernovae, provided that cosmic
evolution proceeds via the regime dominated by the density of a
free scalar field $Q$ \cite{ppgc,039}. Thus, the identification of
conformal coordinates and variables used in observational
cosmology and in quantum field theory with measured quantities is
a first condition under which the origin of matter can be
explained by its cosmological creation from a vacuum. This is
possible within a conformally invariant unified theory, where the
Planck mass, which is an absolute quantity in the general theory
of relativity, becomes an ordinary present-day value of the
dilaton and where the Planck era loses its absolute meaning.

The construction of a stable vacuum of perturbation theory by
eliminating (through the choice of gauge-invariant variables)
unphysical fields whose quantization leads to a negative
normalization of the wave function in this reference frame is a
second condition.

  Finally, the elimination of divergences in
  summing the probabilities of product particles over
  their momenta by thermalizing these particles in the
  region where the Boltzmann
  H-theorem is applicable is a third condition.

Under these conditions, it has been found in the present study
that, in describing the creation of vector bosons from a vacuum in
terms of conformal variables,  one arrives  at the temperature
$(M^2_WH_0)^{1/3}$ $\sim  2.7$K, of cosmic microwave background
radiation as an integral of the motion of the Universe and at the
baryon-antibaryon asymmetry of the Universe with the
superweak-interaction coupling constant $X_{\rm CP}=\frac{n_{\rm
b}}{n_\gamma}$ and the baryon density
$\Omega_b=\frac{\alpha_{\mbox{\tiny QED}}}{\sin^2 \theta_{(W)}}
\sim 0.03$, these results being in satisfactory agreement with the
corresponding observed values and being compatible with the most
recent data on supernovae and nucleosynthesis.

\medskip

\centerline {\it ACKNOWLEDGMENTS}

\medskip

We are grateful to B.A. Arbuzov, B.M. Barbashov, A.V. Efremov,
V.B. Priezzhev, and P. Flin for stimulating discussions. We are
also indebted to the participants of the seminar held at the
Shternberg State Astronomical Institute and dedicated to the
memory of A.L. Zel'manov, especially to M.V. Sazhin and A.A.
Starobinsky, for discussions on the problem of choosing the units
of measurements in cosmology and on the cosmological creation of
massive bosons.

\begin{center}
\it REFERENCES
\end{center}

\vspace{-2cm }

\newpage

\end{document}